\documentclass[journal=jacsat,manuscript=article]{achemso}

\usepackage[version=3]{mhchem} 
\usepackage{siunitx}
\usepackage{graphicx}
\usepackage{gensymb}
\usepackage{bm}
\usepackage{setspace}
\usepackage{xkeyval}
\usepackage[utf8]{inputenc}
\usepackage{amsmath}
\usepackage{array}
\usepackage{rotating}
\usepackage{float}

\title{Influence of Contacts and Applied Voltage on a Structure of a Single GaN Nanowire}

\author{Sergey~Lazarev}
\email{dr.s.lazarev@gmail.com}
\affiliation{Deutsches Elektronen-Synchrotron DESY, Notkestra\ss{e} 85, 22607 Hamburg, Germany}
\alsoaffiliation{National Research Tomsk Polytechnic University (TPU), Lenin Avenue 30, 634050 Tomsk, Russia}

\author{Luca~Gelisio}
\affiliation{Deutsches Elektronen-Synchrotron DESY, Notkestra\ss{e} 85, 22607 Hamburg, Germany}

\author{Young~Yong~Kim}
\affiliation{Deutsches Elektronen-Synchrotron DESY, Notkestra\ss{e} 85, 22607 Hamburg, Germany}

\author{Bi~Zhaoxia}
\affiliation{NanoLund, Department of Physics, Lund University, P.O. Box 118, SE-221 00 Lund, Sweden}

\author{Ali~Nowzari}
\affiliation{NanoLund, Department of Physics, Lund University, P.O. Box 118, SE-221 00 Lund, Sweden}

\author{Ivan~A.~Zaluzhnyy}
\affiliation{Deutsches Elektronen-Synchrotron DESY, Notkestra\ss{e} 85, 22607 Hamburg, Germany}

\author{Ruslan~Khubbutdinov}
\affiliation{Deutsches Elektronen-Synchrotron DESY, Notkestra\ss{e} 85, 22607 Hamburg, Germany}
\alsoaffiliation{National Research Nuclear University MEPhI (Moscow Engineering Physics Institute), Kashirskoe shosse 31, 115409 Moscow, Russia}

\author{Dmitry~Dzhigaev}
\affiliation{Deutsches Elektronen-Synchrotron DESY, Notkestra\ss{e} 85, 22607 Hamburg, Germany}

\author{Arno~Jeromin}
\affiliation{Deutsches Elektronen-Synchrotron DESY, Notkestra\ss{e} 85, 22607 Hamburg, Germany}	

\author{Thomas~Keller} 
\affiliation{Deutsches Elektronen-Synchrotron DESY, Notkestra\ss{e} 85, 22607 Hamburg, Germany}

\author{Michael~Sprung}
\affiliation{Deutsches Elektronen-Synchrotron DESY, Notkestra\ss{e} 85, 22607 Hamburg, Germany}

\author{Anders~Mikkelsen}
\affiliation{NanoLund, Department of Physics, Lund University, P.O. Box 118, SE-221 00 Lund, Sweden}

\author{Lars~Samuelson}
\affiliation{NanoLund, Department of Physics, Lund University, P.O. Box 118, SE-221 00 Lund, Sweden}

\author{Ivan~A.~Vartanyants}%
\email{ivan.vartaniants@desy.de}
\affiliation{Deutsches Elektronen-Synchrotron DESY, Notkestra\ss{e} 85, 22607 Hamburg, Germany}	
\alsoaffiliation{National Research Nuclear University MEPhI (Moscow Engineering Physics Institute), Kashirskoe shosse 31, 115409 Moscow, Russia}

\begin{document}

\begin{abstract}
Semiconductor nanowires (NWs) have a broad range of applications for nano- and optoelectronics. The strain field of gallium nitride (GaN) NWs could be significantly changed when contacts are applied to them to form a final device, especially considering the piezoelectric properties of GaN. Investigation of influence of the metallic contacts on the structure of the NWs is of high importance for their applications in real devices. We have studied a series of different type of contacts and influence of the applied voltage bias on the contacted GaN NWs with the length of about 3 to 4 micrometers and with two different diameters of 200~nm and 350~nm. It was demonstrated that the NWs with the diameter of 200~nm are bend already by the interaction with the substrate. For all GaN NWs, significant structural changes were revealed after the contacts deposition. The results of our research may contribute to the future optoelectronic applications of the GaN nanowires.
\end{abstract}

\section{Introduction}

Semiconductor gallium nitride (GaN) nanowires (NWs) are promising candidates for a number of applications such as light emitting diodes (LED), transistors, single photon sources, low-cost solar cells, and other devices~\cite{Li2017, WeiLu2014NW, Feng2017, Dasgupta2014, Kuykendall2007, Huang2010}.
	Contacting process of the nanostructures in a final optoelectronic device may lead to significant changes in their structure and properties.
In addition, small dimensions and high surface to volume ratio of the NWs result in higher piezoelectric effects, which influence the electron-hole recombination.

The nanostructures are commonly analyzed by the laboratory table-top equipment employing different techniques such as scanning electron microscopy (SEM), atomic force microscopy (AFM), transmission electron microscopy (TEM), laboratory X-ray diffraction (XRD), \textit{etc}.
	In most of the cases, these methods have limitations, which do not allow investigation of the local structure of a single object along the whole NW with high resolution.
Modern $3^{rd}$ generation synchrotron facilities allow investigation of individual nanostructures with intense, highly-coherent, focused down to sub-micrometer scale X-ray beams.
	Various coherent X-ray diffraction techniques such as coherent diffraction imaging (CDI), Bragg CDI, and ptychography may provide an important information about the three-dimensional (3D) strain field and deformation of a single nanostructure with a high spatial resolution~\cite{Newton2010, godard2011three, hruszkewycz2012quantitative, dzhigaev2016bragg, Dzhigaev2017}.

In our previous study~\cite{Lazarev2018}, we have demonstrated that deposition of the Au contacts as well as the applied voltage bias deform single GaN nanowires leading to their bending.
    It was also demonstrated that the arching of the nanostructures may reach the critical value, when the NW is broken.
In our current study, we will further investigate the influence of the various type of contacts and applied voltage on the strain field and structure of the NWs with the two different diameters.

\section{Experiment}

\subsection{Samples}
	
The samples with GaN NWs were prepared at the NanoLund Laboratory at Lund University, Sweden.
	The GaN NWs with the lengths from 3 $\mu$m to 4 $\mu$m and two different diameters of 200~nm and 350~\si{nm} were grown by selective area metal-organic vapor phase epitaxy (MOVPE), equipped with a 3x2 close-coupled showerhead reactor \cite{hersee2006controlled}.
First, a 1~\si{\micro\meter} thick [0001]-oriented GaN template layer was grown on top of a Si (111) substrate, on which a 30 nm thick $SiN_x$ film was deposited as a growth mask by low-pressure chemical vapor deposition (CVD).
	Further, hexagonal arrays of openings were defined in the $SiN_x$ mask by e-beam lithography and reactive ion etching with the opening diameter of about 100 nm and the pitch of 1~$\mu${m}.
The as-patterned templates were then loaded into the MOVPE reactor to grow the GaN nanowires.
	Due to the growth selectivity, the GaN growth only took place from the GaN surface exposed in the $SiN_x$ openings.
The continuously supplied flows of $NH_3$ and triethylgallium were used to synthesize GaN nanowires.
	A low V/III ratio had to be used in order to achieve the nanowire geometry with vertical m-plane side facets.
The triethylgallium flow was 19~$\mu$mol/min and the growth temperature was $1042^\circ$C.
	The as-grown GaN nanowires, having a hexagonal cross-section, were about 200 nm thick and about 3.5~$\mu${m} long.
In order to further increase the diameter of GaN nanowires, a GaN shell was grown around the GaN nanowires.
	This was achieved by increasing the V/III ratio, which enhanced the GaN growth on the m-plane side facets through forming low-growth-rate facets ${10\bar{1}1}$ at the nanowire tip.
By controlling the shell growth time, GaN nanowires with a diameter of 350 nm were obtained.
    From our previous study~\cite{Lazarev2018}, we know that the structure of GaN NWs is pure wurtzite almost without dislocations and stacking faults.

\begin{figure}[H]
		\includegraphics[width=10.5 cm]{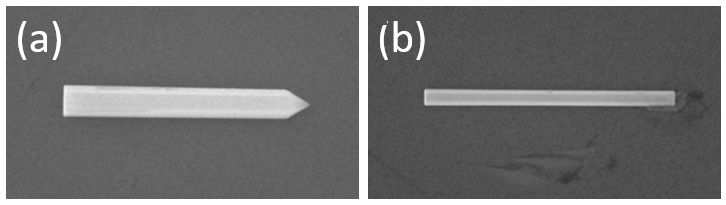}
		\caption{
		SEM images of the free-lying GaN NWs with the diameters of 350~nm (a) and 200~nm (b).
\label{Fig_SEM_contacts_free}}
\end{figure}

After the growth, GaN NWs were removed from the original substrate and deposited on a Si (111) chip with a 100~nm thermally grown $SiO_2$ layer on top, as an insulating layer.
    The SEM images of the free-lying NWs are presented in Fig.~\ref{Fig_SEM_contacts_free}.
    The SEM images of the NWs presented in this work were obtained in NanoLund and DESY NanoLab~\cite{Stierle2016}.

\subsection{Contacts}

\begin{figure}[H]
		\includegraphics[width=10.5 cm]{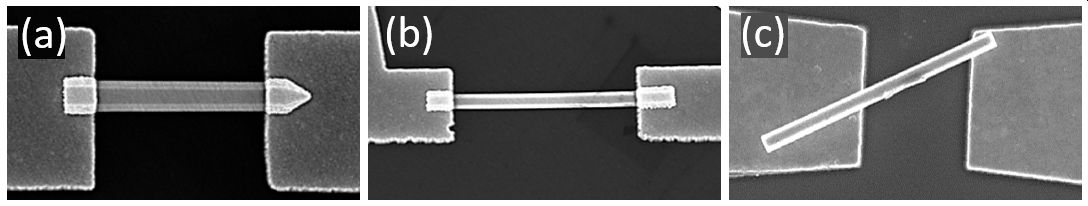}
		\caption{
		SEM images of the GaN NWs with the diameters of 350 nm (a) and 200 nm (b) contacted by the Au electrodes deposited above the NWs.
		    (c) The NW with the diameter of 200 nm contacted on the top of the Au electrodes by melting procedure (see text for details).
\label{Fig_SEM_contacts}}
\end{figure}

After the growth and deposition, Au contacts of two different types were used to connect the GaN NWs in order to study the influence of the applied voltage on their structure.
	The first contacting approach was similar to our previous work~\cite{Lazarev2018}.
The position and orientation of the NWs were random with respect to the main contacts on the substrate after their deposition.
	We developed a sample holder, which provided the electrical connection to the selected single NWs.
The free-lying GaN NWs were connected from two sides to the power supply via pads and solid wires using different contacting methods.
        The 220~nm thick metallic contacts were deposited by electron beam lithography and thermal evaporation of Ti and Au.
	First, a 20~nm thick layer of Ti was deposited on $SiO_2$ and GaN NW to provide good adhesion between Au and GaN, and then 200~nm thick layer of Au on top of Ti layer.
The SEM images of the 350~nm and 200~nm NWs contacted using this method are shown in Fig.~\ref{Fig_SEM_contacts}(a,b).
	In comparison to our previous experiment~\cite{Lazarev2018}, these Au contacts had larger width and were thicker.
The contacts were covering a larger part of the NWs at the bottom, while the tip of the NWs was covered less.
	These variations in contacting geometry define the possible difference in properties of the electronic system "GaN NW + Au contacts".

The second type of contacts was manufactured differently.
	The Au contacts were deposited on the $SiO_2$ layer using the similar lithography technique from two sides leaving a gap between them.
Further, GaN NWs were positioned on the top of these contacts and were slightly melted into the contacts by the increased temperature of 400 C for 20 min.
	The SEM image of this type of contacts is shown in Fig.~\ref{Fig_SEM_contacts}(c).

We performed electrical measurements before the experiment.
	The resistance of the system "GaN NW + Au contacts" was about $10^{12}$~Ohm.
Therefore, the current through the NWs was relatively low, and additional cooling of the sample was not necessary.

\subsection{Experimental setup}		

\begin{figure}[H]
		\includegraphics[width=10.5 cm]{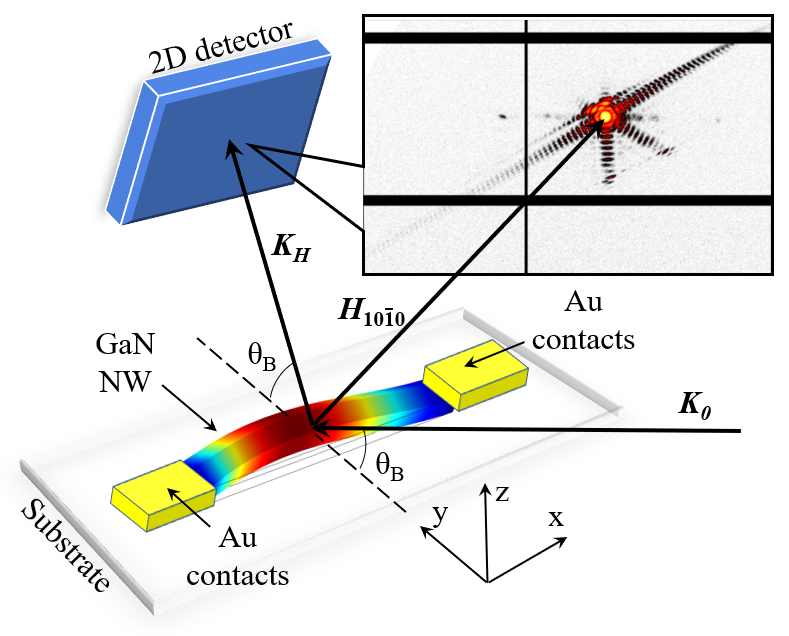}
		\caption{
			The experimental setup showing the incoming and diffracted beams with the wave vectors \textbf{\textit{K}}$_0$ and \textbf{\textit{K}}$_H$ and a
            reciprocal lattice vector \textbf{\textit{H}}$_{10\bar{1}0}$ of the NW that is perpendicular to the Si substrate.
		    The sample is rotated around the Bragg angle $\theta_B$ of $10\bar{1}0$ GaN reflection.
		    The rectangular coordinate system has z-axis perpendicular to the substrate, x-axis along, and y-axis perpendicular to the NW.
            In the inset, the intensity distribution of a typical free-lying NW recorded by a 2D detector is shown.
\label{NW_Experimental_Setup}}
\end{figure}

The nanostructures were investigated at the coherence applications beamline P10 at the PETRA III synchrotron facility (DESY, Hamburg, Germany).
	The geometry of the experiment is shown in Fig.~\ref{NW_Experimental_Setup}.
The measurements were performed at the six-circle diffractometer equipped with a two-dimensional (2D) X-ray pixel detector EIGER X 4M positioned 1.74~m downstream from the sample in Bragg geometry.
	The detector had 2070~x~2167~pixels with the pixel size of 75~x~75~$\mu m^2$.
To reduce air scattering, an evacuated flight tube was mounted between the sample and detector.

The X-ray beam with the photon energy of 9.04~keV and flux of about $10^{11}$~ph/s was focused at the sample using compound refractive lenses (CRLs).
	Characterization of the focus was performed by the knife edge scan at the sample position.
The X-ray beam was about 3~x~2~$\mu m^2$ (V x H) in size at full width at half maximum (FWHM).
	The sample with GaN NWs was mounted on the diffractometer by a specially manufactured  adapter with wires connected to a voltage power supply (see for details Ref.~\cite{Lazarev2018}).
During the experiment, $10\bar{1}0$ Bragg reflection of GaN was measured employing the incoming and diffracted X-ray beams with the wave vectors \textbf{\textit{K}}$_0$ and \textbf{\textit{K}}$_H$ and a reciprocal lattice vector \textbf{\textit{H}}$_{10\bar{1}0}$ of the GaN NW being normal to the substrate (see Fig.~\ref{NW_Experimental_Setup}).
    The sample was positioned at a Bragg angle $\theta_B$ of $10\bar{1}0$ GaN reflection and the detector was placed at $2\theta_B$ with respect to the incident X-ray beam.
The rectangular coordinate system was chosen to have the z-axis perpendicular to the substrate, x-axis along, and y-axis perpendicular to the NW (see Fig.~\ref{NW_Experimental_Setup}).
The 3D coherent intensity distribution around each diffraction peak was measured by recording the scattered X-rays from the specimen by the 2D detector at different incidence angles in a range of about $\pm0.4\degree$ around the Bragg angle with 160 angular steps and 5 seconds of exposure time.

\section{Results}

\subsection{Free-lying nanowires}

\begin{figure}[H]
		\includegraphics[width=10.5 cm]{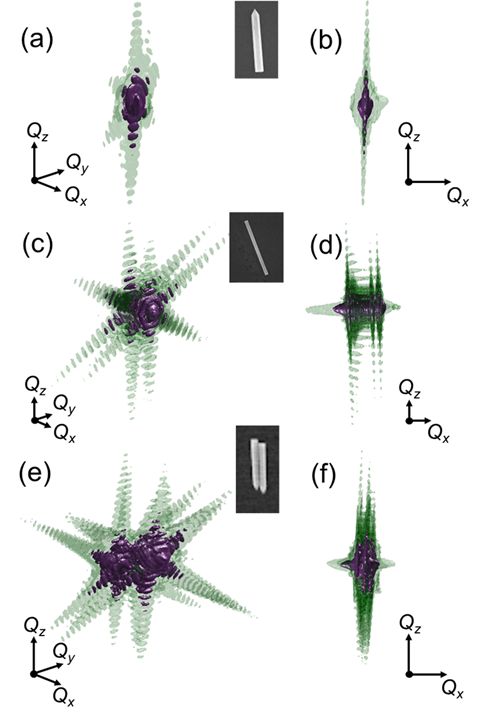}
		\caption{
			The intensity distribution around $10\bar{1}0$ GaN Bragg reflection of free-lying GaN NWs shown from two different views to emphasize the
            diffraction pattern details.
			(a-d) Single GaN NWs with the diameters of 350~nm (a, b) and 200~nm (c, d).
			The thinner 200~nm NW reveals a double Bragg peak structure due to its bending on the substrate (see text for details).
			(e, f) For a comparison, the intensity distribution of two NWs with the diameters of 350~nm lying close to each other form well separated Bragg
            peaks in reciprocal space.
			To enhance the diffraction pattern details, the intensity distributions are represented by two different iso-surfaces.
			The length of coordinate arrows corresponds to 0.1 nm$^{-1}$.
			The intensity is normalized to the maximum and the iso-surface values are $10^{-3}$ and $10^{-4.5}$ (a,b), $10^{-0.5}$ and $10^{-1}$ (c,d),
            $10^{-0.5}$ and $10^{-1}$ (e,f).
\label{Fig_Free_NWs}}
\end{figure}

First, we have investigated a series of free-lying NWs without contacting electrodes.
	The 3D distribution of coherently scattered intensity in the vicinity of a $10\bar{1}0$ GaN Bragg peak of the free-lying NWs in the laboratory frame is shown in Fig.~\ref{Fig_Free_NWs}.
The intensity distribution of a free-lying NW of 350~nm in size is presented in Fig.~\ref{Fig_Free_NWs}(a,b) and demonstrates well pronounced hexagonal symmetry with the fringes originating from the opposite facets of the NW.
    There is only a single star-shaped Bragg peak distribution in reciprocal space (see Fig.~\ref{Fig_Free_NWs}(b)), which indicates that this NW does not have a significant structural change and may be considered as a deformation-free NW.
Several other investigated free-lying NWs of the same size demonstrated similar structure of the Bragg peak distribution in reciprocal space.

Further, a single, free-lying GaN NW with a smaller diameter of 200~nm was studied.
    Its intensity distribution is shown from two different directions in Figs.~\ref{Fig_Free_NWs}(c,d).
Interestingly, the diffraction pattern reveals the "double-star" structure of the Bragg peak distribution (see Fig.~\ref{Fig_Free_NWs}(d)).
    Two additionally studied 200~nm free-lying GaN NWs demonstrated similar double structure of the Bragg peak intensity distribution, which is due to the bending of the NWs~\cite{Lazarev2018}.

    For comparison, a diffraction pattern of $10\bar{1}0$ GaN Bragg peak of two close lying NWs with diameters of 350~nm is presented in Figs.~\ref{Fig_Free_NWs}(e,f).
In this case of 350~nm thick NWs, two well-separated hexagonal-star Bragg peaks are distinguishable in reciprocal space.

Therefore, it is possible to conclude that the free-lying GaN NWs with the diameter of 200~nm are already bent in contrast to 350~nm in diameter NWs.
	This effect makes investigation of the NWs with the diameters smaller than 350~nm more complicated.

\subsection{Influence of contacts on the NWs structure}

\begin{figure}[H]
		\includegraphics[width=10.5 cm]{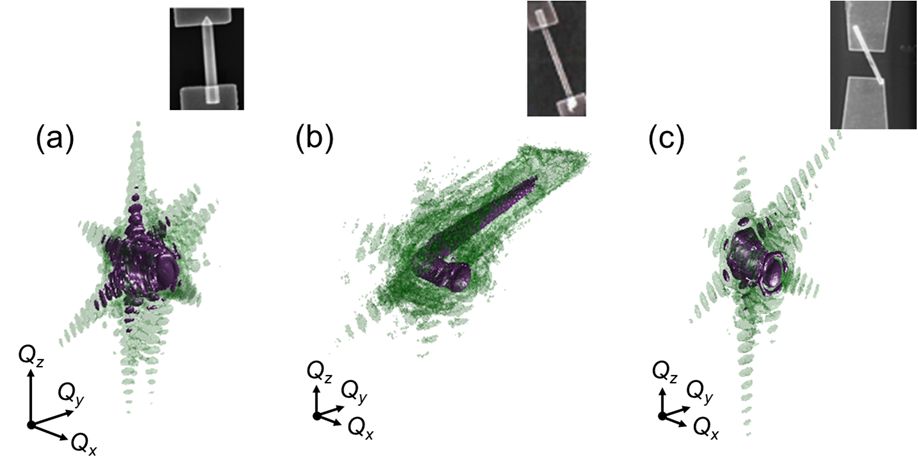}
		\caption{
			Comparison of the intensity distributions around $10\bar{1}0$ GaN NWs Bragg peak for the different types of Au contacts.
			(a, b) The contacts deposited on the top of the nanowires have the diameters of 350~nm (a) and 200~nm (b).
			(c) The intensity distribution for a NW with the diameter of 200~nm contacted on the top of the Au electrodes by melting procedure.
			For all types of the contacts, a significant influence on the scatted X-ray intensity due to the induced strain during the contacting process
            could be seen.
			To enhance the diffraction pattern details, the intensity distributions are represented by two different iso-surfaces.
		    The length of coordinate arrows corresponds to 0.1 nm$^{-1}$.
		    The intensity is normalized to the maximum and the iso-surface values are $10^{-3.5}$ and $10^{-4.5}$ (a), $10^{-1.5}$ and $10^{-2.85}$ (b),
            $10^{-2.5}$ and $10^{-3.5}$ (c).
\label{Fig_Before_After_contacts}}
\end{figure}

The 3D intensity distribution around $10\bar{1}0$ GaN Bragg peak for the contacted nanowires is shown in Fig.~\ref{Fig_Before_After_contacts}.
    It is well seen in Fig.~\ref{Fig_Before_After_contacts}(a) that the Bragg peak of the contacted 350~nm thick GaN NW has obtained an additional Bragg peak due to the induced strain from the deposited contacts.

The free-lying 200~nm thin GaN NWs were already bent before contacts deposition, but the bending angle was increased by 2 degrees after the contacts were applied (see Fig.~\ref{Fig_Before_After_contacts}(b)).
	The nanowire fixation and bending mechanism were different for 350~nm and the 200~nm GaN NWs as can be seen from comparison of corresponding diffraction patterns (see Fig.~\ref{Fig_Before_After_contacts}(a,b)).

Further, influence of the second type of Au contacts on the GaN NWs was studied
(see Fig.~\ref{Fig_SEM_contacts}(c)).
    The corresponding distribution of intensity in reciprocal space around $10\bar{1}0$ GaN Bragg peak of a contacted 200~nm GaN NW is presented in Fig.~\ref{Fig_Before_After_contacts}(c), and shows the bending effect.
Finally, we can conclude that all investigated types of contacts had a similar bending effect on the GaN NWs.

\subsection{In-operando studies of GaN NWs}			
		
\begin{figure}[H]
		\includegraphics[width=10.5 cm]{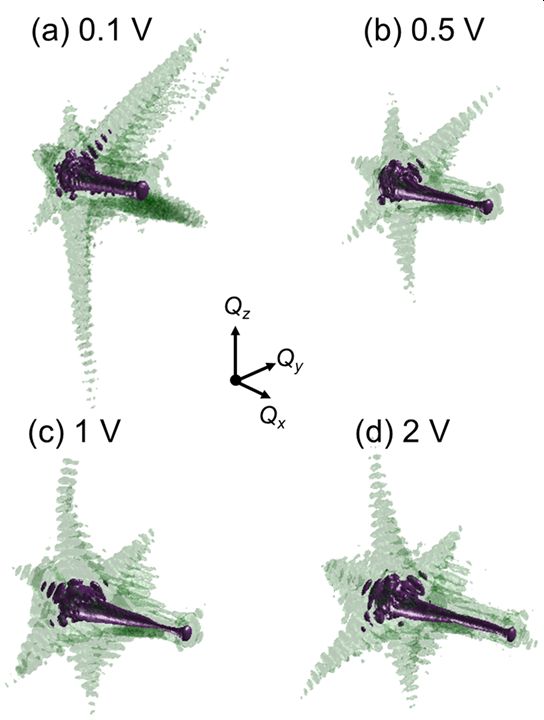}
		\caption{
			Evolution of intensity distribution around $10\bar{1}0$ GaN Bragg reflection of a contacted GaN NW with the diameter of 350~nm. 	
		    The values of the applied voltage bias: 0.1~V (a), 0.5~V (b), 1~V (c), 2~V (d).
			To enhance the diffraction pattern details, the intensity distributions are represented by two different iso-surfaces.
		    The length of coordinate arrows corresponds to 0.1 nm$^{-1}$.
		    The intensity is normalized to the maximum and the iso-surface values are $10^{-3.4}$ and $10^{-4}$.
\label{Fig_Au_voltage_NW_4}}
\end{figure}

To study properties of the GaN NWs with the diameter of 350~nm, we have applied voltage bias to their ends and followed the evolution of the $10\bar{1}0$ GaN Bragg diffraction pattern.
	The Bragg peak intensity distribution as a function of applied voltage is shown in Fig.~\ref{Fig_Au_voltage_NW_4}.
It is clearly visible in the figure that the distance between the Bragg peaks increases with the voltage causing an additional redistribution of the coherently scattered radiation around the Bragg peaks.
    As was already discussed above, this deformation of the Bragg peak is attributed to the GaN NW bending due to its expansion caused by the piezo effect under the applied voltage.
The scheme of the bent NW and its intensity distribution in reciprocal space is shown in Fig.~\ref{Fig_Voltage_Angle_Au}(a).
	The gap between two peaks increased with the applied voltage till the maximum elongation at two volts (see Fig.~\ref{Fig_Au_voltage_NW_4}(d)).
After this critical value, the applied voltage bias was increased to five volts, and the NW was destroyed.
	Therefore, we assumed that the breakdown voltage of the system "GaN NW + Au contacts" was between two and five volts.

Investigation of the applied voltage to the GaN NWs with a diameter of 200~nm was not successful due to the unstable contacts between the NWs and the metallic contacts.
	The 200~nm GaN NW positioned on the top of the second type of Au contacts was lost after the first applied voltage bias of 0.5 volts.
In this situation, one can only conclude that this type of NW fixation is much weaker than the variant with the Au contacts on the top of the NW.

\section{Discussion}

\begin{figure}[H]
		\includegraphics[width=10.5 cm]{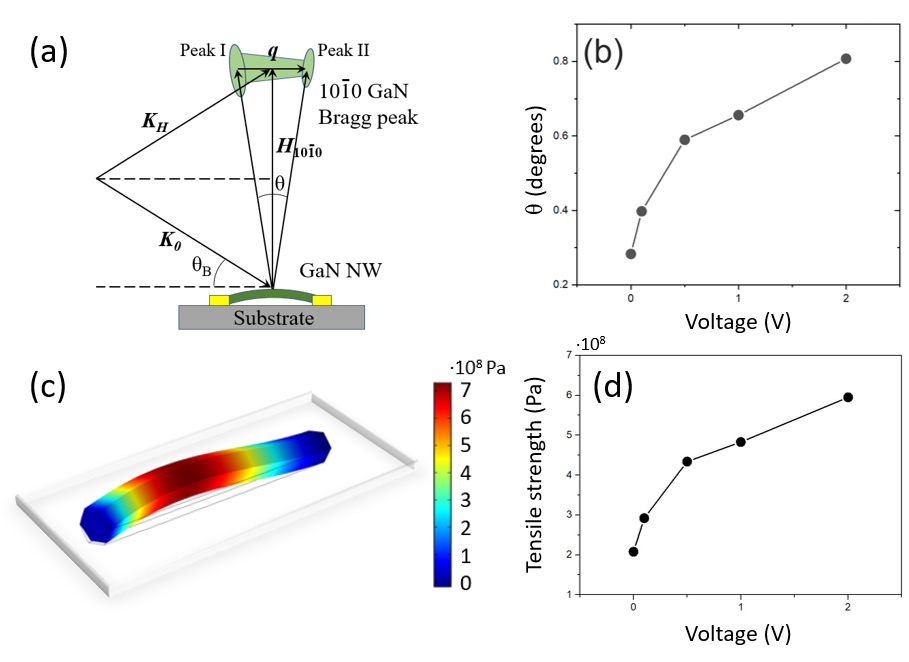}
		\caption{
		(a) Scheme of reciprocal space X-ray intensity distribution formation.
		(b) Bending angle $\theta$ of the NW with the size of 350 nm as a function of the applied voltage bias.
		(c) 3D FEM model of the bent GaN NW.
		(d) Maximum tensile strength in the NW under voltage bias.
		The stress in the NW is increased from $2\cdot10^8$Pa up to the ultimate tensile strength of $6\cdot10^8$Pa, at which the nanowire was broken.
		\label{Fig_Voltage_Angle_Au}}
\end{figure}

    To analyze the bending effect of the GaN NWs, a model of their bending was developed.
A scheme of the intensity distribution in reciprocal space is shown in Fig.~\ref{Fig_Voltage_Angle_Au}(a), and explains relation between the bending angle $\theta = \arctan(q/\textbf{$H_{10\bar{1}0}$})$ of the NW and the elongation $q$ of the gap between two star-like Bragg peaks with the known length of the scattering vector \textbf{\textit{H}}$_{10\bar{1}0}$.	
    Additionally, a finite element method (FEM) model of the contacted nanostructure, developed in our previous work~\cite{Lazarev2018}, was used to correlate $\theta$ with the inner stain of the NW.

For the free-lying NWs with the diameter of 200~nm, an average bending angle of 0.45 degrees was calculated from the distances $q$ between two peaks in reciprocal space (see Fig.~\ref{Fig_Voltage_Angle_Au}(a).
    Additionally, the FEM model was used to obtain the maximum tensile strength in the NWs, which was found to be about 0.34~GPa.

The bending angle of the contacted 350 nm GaN NWs was found to be 0.28 degrees and corresponds to $2\cdot10^8$~Pa tensile strength obtained from the FEM model.
    In our previous experiment, the GaN NW was bent to 0.4 degrees after the Au contacts deposition with the tensile strength of $3\cdot10^8$~Pa, which supports the current findings.
For the second type Au contacts, Fig.~\ref{Fig_Before_After_contacts}(c) shows an elongated Bragg peak due to the NW bending.
    The angle of 0.29 degrees was obtained in the case of this type of Au contacts.

Dependence of the bending angle $\theta = \arctan(q/\textbf{$H_{10\bar{1}0}$})$  between the two peaks of the $10\bar{1}0$ GaN Bragg reflection for the GaN NW under in-operando investigation is given in Fig.~\ref{Fig_Voltage_Angle_Au}(b) as a function of the voltage bias.
        The 3D view of the FEM model is presented in Fig.~\ref{Fig_Voltage_Angle_Au}(c).
	From the FEM simulations, the maximum tensile strength of the most top part of the nanowire was obtained for different bending angles  measured in the experiment.
Dependence of the maximum tensile strength as a function of applied voltage bias is presented in Fig.~\ref{Fig_Voltage_Angle_Au}(d).
	The strain in the GaN NW increased from $2\cdot10^8$Pa up to $6\cdot10^8$Pa, at which the nanostructure was broken.
In general, the strain values are in a good agreement with the dependence observed in our previous study, but the voltage values corresponding to the same bending angles are five times smaller.
	We attribute this discrepancy to the different geometric parameters of the contacts, which were thicker this time and had different overlap with the NWs.
The ultimate tensile strength at the breakdown voltage could be estimated with rather large error from the extrapolation of the obtained tensile strength dependence, but will be in a range of the previously obtained value of 1~GPa.

\section{Summary and outlook}

In our study, a strong influence of the different types of metallic contacts on the strain state of GaN NWs with diameters of 200~nm and 350~nm was demonstrated using coherent X-ray diffraction of the $10\bar{1}0$ GaN Bragg peak.
    Employing the developed FEM model, the maximum tensile strength in the nanowires was calculated for the different bending states induced due to the deposited Au contacts.

For the NWs with 350~nm in diameter, the maximum tensile strength raised to $2\cdot10^8$Pa after the contacts deposition.
	Further, the influence of the applied voltage bias on the strain field evolution until the breakdown of the nanostructures was investigated.
The strain in the GaN NWs increased up to the ultimate tensile strength of $6\cdot10^8$Pa corresponding to two volts of applied bias, at which the nanowire was broken.
    This ultimate tensile strength was in a good agreement with out previous work, while the breakdown voltage deviation was explained by the slightly different geometry of the contacts and their possibly different breakdown behavior.

In the case of the NWs with the diameter of 200~nm, it was demonstrated that they bend already by the interaction with the substrate.
    The applied Au contacts disturbed the scattered X-ray intensity distribution of the Bragg peak even more.
Additionally, the contacts fixation of this type of NWs was less stable, which made their investigation with the current diameter to length ratio rather challenging.
    For our future experiments, the length of the NWs with diameters below 350~nm is planned to be shorter than 2~$\mu$m.

An additional type of fixation with the NWs positioned on the top of the Au contacts demonstrated for the GaN NWs with the diameter of 200~nm a much weaker "contact-nanostructure" fixation.
    The GaN NW was disconnected already after the first applied voltage bias of 0.5 volts.
Due to these reasons, the in-operando study of the nanowires with this type contacts was not successful.

We expect that the results of our work will contribute to the development and manufacture of the devices based on the nanostructures and may help the high-technology industry development.

\vspace{6pt}



\section{Authors contribution}

Conceptualization and methodology, S.L. and I.A.V.; sample preparation, B.Z., A.N., A.M., and L.S.; sample characterization, A.J. and T.F.K.; X-ray experiment, S.L., Y.Y.K., L.G., I.A.Z., R.K., D.D., M.S., and I.A.V.; data analysis, S.L., Y.Y.K., and D.D.; writing---review and editing, S.L. and I.A.V. All authors have read and agreed to the published version of the manuscript.

\section{Funding}
This research was funded by the Helmholtz Associations Initiative Networking Fund (Grant No. HRSF-0002) and the Russian Science Foundation (Grant No. 18–41-06001); Sergey Lazarev was funded by the Competitiveness Enhancement Program Grant of Tomsk Polytechnic University and the Governmental program “Science” project no. FSWW-2020-0014.

\section{Acknowledgments}
We acknowledge DESY (Hamburg, Germany), a member of the Helmholtz Association HGF, for the provision of experimental facilities. Parts of this research were carried out at PETRA III and DESY NanoLab and we would like to thank the beamline staff for assistance in using coherence applications beamline P10. The authors are thankful to E. Weckert for the support of the project and also acknowledge a careful reading of the manuscript by Z. Ren.

\section{Conflicts of interest}
The authors declare no conflict of interest.









\bibliography{bibliography}

\providecommand{\latin}[1]{#1}
\makeatletter
\providecommand{\doi}
  {\begingroup\let\do\@makeother\dospecials
  \catcode`\{=1 \catcode`\}=2 \doi@aux}
\providecommand{\doi@aux}[1]{\endgroup\texttt{#1}}
\makeatother
\providecommand*\mcitethebibliography{\thebibliography}
\csname @ifundefined\endcsname{endmcitethebibliography}
  {\let\endmcitethebibliography\endthebibliography}{}
\begin{mcitethebibliography}{15}
\providecommand*\natexlab[1]{#1}
\providecommand*\mciteSetBstSublistMode[1]{}
\providecommand*\mciteSetBstMaxWidthForm[2]{}
\providecommand*\mciteBstWouldAddEndPuncttrue
  {\def\EndOfBibitem{\unskip.}}
\providecommand*\mciteBstWouldAddEndPunctfalse
  {\let\EndOfBibitem\relax}
\providecommand*\mciteSetBstMidEndSepPunct[3]{}
\providecommand*\mciteSetBstSublistLabelBeginEnd[3]{}
\providecommand*\EndOfBibitem{}
\mciteSetBstSublistMode{f}
\mciteSetBstMaxWidthForm{subitem}{(\alph{mcitesubitemcount})}
\mciteSetBstSublistLabelBeginEnd
  {\mcitemaxwidthsubitemform\space}
  {\relax}
  {\relax}

\bibitem[Li \latin{et~al.}(2017)Li, Wright, Liu, Lu, Figiel, Leung, Chow,
  Brener, Koleske, Luk, Feezell, Brueck, and Wang]{Li2017}
Li,~C.; Wright,~J.~B.; Liu,~S.; Lu,~P.; Figiel,~J.~J.; Leung,~B.; Chow,~W.~W.;
  Brener,~I.; Koleske,~D.~D.; Luk,~T.-S.; Feezell,~D.~F.; Brueck,~S. R.~J.;
  Wang,~G.~T. {Nonpolar InGaN/GaN Core–Shell Single Nanowire Lasers}.
  \emph{Nano Lett.} \textbf{2017}, \emph{17}, 1049--1055\relax
\mciteBstWouldAddEndPuncttrue
\mciteSetBstMidEndSepPunct{\mcitedefaultmidpunct}
{\mcitedefaultendpunct}{\mcitedefaultseppunct}\relax
\EndOfBibitem
\bibitem[Lu(2014)]{WeiLu2014NW}
Lu,~W. \emph{{Semiconductor Nanowires: From Next-Generation Electronics to
  Sustainable Energy}}; The Royal Society of Chemistry, 2014\relax
\mciteBstWouldAddEndPuncttrue
\mciteSetBstMidEndSepPunct{\mcitedefaultmidpunct}
{\mcitedefaultendpunct}{\mcitedefaultseppunct}\relax
\EndOfBibitem
\bibitem[Feng(2017)]{Feng2017}
Feng,~Z.~C. \emph{{III-Nitride Materials, Devices and Nano-Structures}}; World
  Scientific Publishing Company Pte Limited, 2017\relax
\mciteBstWouldAddEndPuncttrue
\mciteSetBstMidEndSepPunct{\mcitedefaultmidpunct}
{\mcitedefaultendpunct}{\mcitedefaultseppunct}\relax
\EndOfBibitem
\bibitem[{Dasgupta, N. P. and Sun, J. and Liu, C. and Brittman, S. and Andrews,
  S. C. and Lim, J. and Gao, H. and Yan, R. and Yang, P.}(2014)]{Dasgupta2014}
{Dasgupta, N. P. and Sun, J. and Liu, C. and Brittman, S. and Andrews, S. C.
  and Lim, J. and Gao, H. and Yan, R. and Yang, P.}, {25th Anniversary article:
  semiconductor nanowires--synthesis, characterization, and applications}.
  \emph{Adv. Mater.} \textbf{2014}, \emph{26}, 2137--2184\relax
\mciteBstWouldAddEndPuncttrue
\mciteSetBstMidEndSepPunct{\mcitedefaultmidpunct}
{\mcitedefaultendpunct}{\mcitedefaultseppunct}\relax
\EndOfBibitem
\bibitem[{Kuykendall, T. and Ulrich, P. and Aloni, S. and Yang,
  P.}(2007)]{Kuykendall2007}
{Kuykendall, T. and Ulrich, P. and Aloni, S. and Yang, P.}, {Complete
  composition tunability of InGaN nanowires using a combinatorial approach}.
  \emph{Nat. Mater.} \textbf{2007}, \emph{6}, 951--956\relax
\mciteBstWouldAddEndPuncttrue
\mciteSetBstMidEndSepPunct{\mcitedefaultmidpunct}
{\mcitedefaultendpunct}{\mcitedefaultseppunct}\relax
\EndOfBibitem
\bibitem[Huang \latin{et~al.}(2010)Huang, Song, Tsai, Lee, Lien, Gao, Hao,
  Chen, and Wang]{Huang2010}
Huang,~C.-T.; Song,~J.; Tsai,~C.-M.; Lee,~W.-F.; Lien,~D.-H.; Gao,~Z.; Hao,~Y.;
  Chen,~L.-J.; Wang,~Z.~L. {Single-InN-Nanowire Nanogenerator with Upto 1 V
  Output Voltage}. \emph{Adv. Mater.} \textbf{2010}, \emph{22},
  4008--4013\relax
\mciteBstWouldAddEndPuncttrue
\mciteSetBstMidEndSepPunct{\mcitedefaultmidpunct}
{\mcitedefaultendpunct}{\mcitedefaultseppunct}\relax
\EndOfBibitem
\bibitem[Newton \latin{et~al.}(2010)Newton, Leake, Harder, and
  Robinson]{Newton2010}
Newton,~M.~C.; Leake,~S.~J.; Harder,~R.; Robinson,~I.~K. Three-dimensional
  imaging of strain in a single ZnO nanorod. \emph{Nat. Mater.} \textbf{2010},
  \emph{9}, 120–124\relax
\mciteBstWouldAddEndPuncttrue
\mciteSetBstMidEndSepPunct{\mcitedefaultmidpunct}
{\mcitedefaultendpunct}{\mcitedefaultseppunct}\relax
\EndOfBibitem
\bibitem[Godard \latin{et~al.}(2011)Godard, Carbone, Allain, Mastropietro,
  Chen, Capello, Diaz, Metzger, Stangl, and Chamard]{godard2011three}
Godard,~P.; Carbone,~G.; Allain,~M.; Mastropietro,~F.; Chen,~G.; Capello,~L.;
  Diaz,~A.; Metzger,~T.~H.; Stangl,~J.; Chamard,~V. {Three-dimensional
  high-resolution quantitative microscopy of extended crystals}. \emph{Nat.
  Commun.} \textbf{2011}, \emph{2}, 568\relax
\mciteBstWouldAddEndPuncttrue
\mciteSetBstMidEndSepPunct{\mcitedefaultmidpunct}
{\mcitedefaultendpunct}{\mcitedefaultseppunct}\relax
\EndOfBibitem
\bibitem[Hruszkewycz \latin{et~al.}(2012)Hruszkewycz, Holt, Murray, Bruley,
  Holt, Tripathi, Shpyrko, McNulty, Highland, and
  Fuoss]{hruszkewycz2012quantitative}
Hruszkewycz,~S.~O.; Holt,~M.~V.; Murray,~C.~E.; Bruley,~J.; Holt,~J.;
  Tripathi,~A.; Shpyrko,~O.~G.; McNulty,~I.; Highland,~M.~J.; Fuoss,~P.~H.
  {Quantitative nanoscale imaging of lattice distortions in epitaxial
  semiconductor heterostructures using nanofocused X-ray Bragg projection
  ptychography}. \emph{Nano Lett.} \textbf{2012}, \emph{12}, 5148--5154\relax
\mciteBstWouldAddEndPuncttrue
\mciteSetBstMidEndSepPunct{\mcitedefaultmidpunct}
{\mcitedefaultendpunct}{\mcitedefaultseppunct}\relax
\EndOfBibitem
\bibitem[Dzhigaev \latin{et~al.}(2016)Dzhigaev, Shabalin, Stankevi{\v{c}},
  Lorenz, Kurta, Seiboth, Wallentin, Singer, Lazarev, Yefanov, Borgstr{\"o}m,
  Strikhanov, Samuelson, Falkenberg, Schroer, Mikkelsen, Feidenhans‘l, and
  Vartanyants]{dzhigaev2016bragg}
Dzhigaev,~D. \latin{et~al.}  {Bragg coherent x-ray diffractive imaging of a
  single indium phosphide nanowire}. \emph{J. Opt.} \textbf{2016}, \emph{18},
  064007\relax
\mciteBstWouldAddEndPuncttrue
\mciteSetBstMidEndSepPunct{\mcitedefaultmidpunct}
{\mcitedefaultendpunct}{\mcitedefaultseppunct}\relax
\EndOfBibitem
\bibitem[Dzhigaev \latin{et~al.}(2017)Dzhigaev, Stankevič, Bi, Lazarev, Rose,
  Shabalin, Reinhardt, Mikkelsen, Samuelson, Falkenberg, Feidenhans’l, and
  Vartanyants]{Dzhigaev2017}
Dzhigaev,~D.; Stankevič,~T.; Bi,~Z.; Lazarev,~S.; Rose,~M.; Shabalin,~A.;
  Reinhardt,~J.; Mikkelsen,~A.; Samuelson,~L.; Falkenberg,~G.;
  Feidenhans’l,~R.; Vartanyants,~I.~A. X-ray Bragg Ptychography on a Single
  InGaN/GaN Core–Shell Nanowire. \emph{ACS Nano} \textbf{2017}, \emph{11},
  6605--6611\relax
\mciteBstWouldAddEndPuncttrue
\mciteSetBstMidEndSepPunct{\mcitedefaultmidpunct}
{\mcitedefaultendpunct}{\mcitedefaultseppunct}\relax
\EndOfBibitem
\bibitem[Lazarev \latin{et~al.}(2018)Lazarev, Dzhigaev, Bi, Nowzari, Kim, Rose,
  Zaluzhnyy, Gorobtsov, Zozulya, Lenrick, Gustafsson, Mikkelsen, Sprung,
  Samuelson, and Vartanyants]{Lazarev2018}
Lazarev,~S.; Dzhigaev,~D.; Bi,~Z.; Nowzari,~A.; Kim,~Y.~Y.; Rose,~M.;
  Zaluzhnyy,~I.~A.; Gorobtsov,~O.~Y.; Zozulya,~A.~V.; Lenrick,~F.;
  Gustafsson,~A.; Mikkelsen,~A.; Sprung,~M.; Samuelson,~L.; Vartanyants,~I.~A.
  Structural Changes in a Single GaN Nanowire under Applied Voltage Bias.
  \emph{Nano Lett.} \textbf{2018}, \emph{18}, 5446--5452\relax
\mciteBstWouldAddEndPuncttrue
\mciteSetBstMidEndSepPunct{\mcitedefaultmidpunct}
{\mcitedefaultendpunct}{\mcitedefaultseppunct}\relax
\EndOfBibitem
\bibitem[Hersee \latin{et~al.}(2006)Hersee, Sun, and
  Wang]{hersee2006controlled}
Hersee,~S.~D.; Sun,~X.; Wang,~X. {The controlled growth of GaN nanowires}.
  \emph{Nano Lett.} \textbf{2006}, \emph{6}, 1808--1811\relax
\mciteBstWouldAddEndPuncttrue
\mciteSetBstMidEndSepPunct{\mcitedefaultmidpunct}
{\mcitedefaultendpunct}{\mcitedefaultseppunct}\relax
\EndOfBibitem
\bibitem[Stierle \latin{et~al.}(2016)Stierle, Keller, Noei, Vonk, and
  Roehlsberger]{Stierle2016}
Stierle,~A.; Keller,~T.~F.; Noei,~H.; Vonk,~V.; Roehlsberger,~R. DESY NanoLab.
  \emph{Journal of large-scale research facilities JLSRF} \textbf{2016},
  \emph{2}, A76\relax
\mciteBstWouldAddEndPuncttrue
\mciteSetBstMidEndSepPunct{\mcitedefaultmidpunct}
{\mcitedefaultendpunct}{\mcitedefaultseppunct}\relax
\EndOfBibitem
\end{mcitethebibliography}

%


\end{document}